\documentclass{phyeauth}
\usepackage{amsmath}
\usepackage{amssymb}
\def\be{\begin{equation}}
\def\ee{\end{equation}}
\include{psfig}
\usepackage{epsfig}

\begin{document}

\begin{frontmatter}

\title{Ground State Entanglement Energetics}

\author{M. B\"uttiker$^1$, A. N. Jordan},

\address{D\'epartement de Physique Th\'eorique, 
Universit\'e de Gen\`eve, CH-1211 Gen\`eve 4, Switzerland.}

\thanks{Corresponding author. \\{\it E-mail adress}:
Markus.Buttiker@physics.unige.ch}

\begin{abstract}
We consider the ground state of simple quantum systems
coupled to an environment. In general the system is
entangled with its environment.
As  a consequence, even at zero temperature, 
the energy of the system is not sharp: a
projective measurement can find the system in an excited state. 
We show that energy fluctuation measurements at zero temperature
provide entanglement information. 
For two-state systems which 
exhibit a persistent current in the ground state,
energy fluctuations and persistent current fluctuations 
are closely related. The harmonic oscillator serves to illustrate 
energy fluctuations in a system with an infinite number of states. 
In addition to the energy distribution we discuss the energy-energy 
time-correlation function in the zero-temperature limit.
\end{abstract}

\end{frontmatter}

\section{Introduction}

A quantum system cooled to zero temperature nevertheless knows about 
its environment since generically the system state and the bath 
state are entangled. The ground state does not factorize into a product
of a system wave function and a bath wave function. Entanglement 
of two subsystems \cite{sch} is often discussed in terms of the strange 
non-local properties it implies for systems that can be spatially 
separated. Here we consider two-level systems, often now called 
qubits, or harmonic oscillators which are coupled to a bath. In such 
a "thermodynamic" setting we can not easily separate the two systems 
and apply a Bell test \cite{bell} to verify the entanglement. 
Nevertheless, systems entangled with reservoir states exhibit a number 
of properties which distinguishes them from systems for which the 
ground state factorizes.

A particularly instructive quantity to consider is the energy of 
the system. First, from a purely theoretical point of view, 
we always have the energy of the system as an observable. We write 
the energy of the total system as 
\be
\hat H = \hat H_s + \hat H_c + \hat H_b
\ee
where $\hat H_s$ is the system,  
$\hat H_c $ the coupling and $\hat H_b$ the bath
energy operator. A second, more important reason for considering the energy is
that at zero temperature fluctuations in energy are a direct indicator 
for system bath entanglement \cite{jb}. In contrast, if the system-bath state 
is not entangled, the system is simply in its lowest energy state. This case
is often viewed as self-evident instead of the generic case addressed here. 

Consider a 
two level system with energies $E_{+}$
and $E_{-}$ and probabilities $p_{\pm}$ to find the system 
in the excited level and in its ground state. The expectation value 
of the energy is 
\begin{equation}
\langle \hat H_s \rangle = 
E_{-} p_{-}+ E_{+}p_{+}
\label{anergy}
\end{equation}
and the energy fluctuation away from its average is 
\begin{equation}
\langle ( \hat H_s - \langle \hat H_s \rangle )^{2} \rangle = 
(E_{+}-E_{-})^{2} p_{-}p_{+} .
\label{energyfl}
\end{equation}
For a moment consider the system to be in thermal equilibrium
coupled to a bath at temperature $T$. 
Later we will only consider the zero-temperature limit. 
Assume for the moment that there is no entanglement.
Then statistical mechanics tells us that 
$p_{+} \propto \exp(-E_{+}/kT)$ and $p_{-} \propto \exp(-E_{-}/kT)$ or 
with proper normalization ($p_{+} + p_{-} =1$)
\begin{equation}
p_{\pm} = \frac{1}{1+ \exp(\pm (E_{+} -E_{-})/kT)}. 
\label{gibbs}
\end{equation}
As the temperature tends to zero, $p_{-}$ tends to one 
($p_{+}$ tends to zero)
and consequently the energy fluctuations Eq. (\ref{energyfl}) tend to zero. 
The energy of the system is a sharp variable in the ground state. 

\begin{figure}[t]
\begin{center}
\psfig{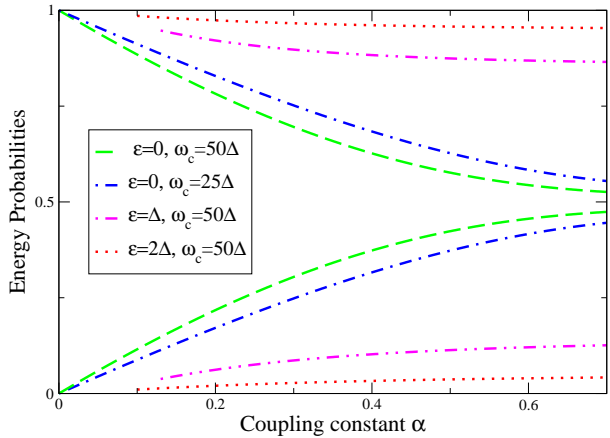}
\caption{Energy probabilities $p_{+}$ and $p_{-}$
for the spin-boson problem as a function of coupling strength 
$\alpha$ and different $\epsilon$. For weak coupling 
to the environment increasing coupling increases the 
the probability to find the qubit in the excited state.  
with increasing 
coupling to the environment it
it is more likely to measure the qubit in the excited state.
After Ref.\protect\cite{jb}}
\label{qubitprobs}
\end{center}
\end{figure}

This simple picture is dramatically modified if the system and 
the bath are entangled. Now statistical mechanics must be applied 
to the entire system (system plus bath plus interaction energy).
The system alone is now determined by a (reduced) density matrix with elements 
$\rho_{\pm,\pm}$ on the diagonal and non-diagonal 
elements $\rho_{\pm,\mp}$. In the energy eigen basis of the two state system, 
$p_{\pm} = \rho_{\pm,\pm}$. While the ground state of the entire 
system is in general a pure state, the density matrix of the system 
alone is in general that of a mixed state. For the simple model 
of a two state system (a spin) coupled to a harmonic oscillator bath 
(the spin-boson problem \cite{rmp,weiss}) an ohmic 
bath leads for weak coupling to a probability \cite{jb}
\begin{equation}
p_{+} =  \alpha \log(\omega_{c} /\Omega ) .
\label{plog}
\end{equation}
Here $\alpha$ is the system bath coupling constant, 
$\hbar \Omega = E_{+} -E_{-}$ is the energy separation of the two levels
and $\omega_c$ is a cut-off of the oscillator spectrum of the bath 
(a Debye frequency). As a consequence, even at zero temperature 
the energy of the system is not sharp but fluctuates according 
to Eq. (\ref{energyfl})
\begin{equation}
\langle ( \hat H_s - \langle \hat H_s \rangle )^{2} \rangle = 
\alpha (\hbar \Omega)^{2} 
\log(\omega_c/\Omega) .
\label{energyflspin}
\end{equation}
We emphasize that this result is not special for a two state system. 
For instance for a harmonic oscillator with frequency 
$\omega_{0}$ coupled to an ohmic bath 
of harmonic oscillators, we find that the 
energy fluctuations\cite{oc,nb} are given by
\begin{equation}
\langle (\hat H_s - \langle \hat H_s \rangle )^{2} \rangle = 
\alpha (\hbar \omega_{0})^{2} 
\log(\omega_c/\omega_{0}) .
\label{energyflosc}
\end{equation}
Thus $\hbar\omega_{0}$ plays a role similar to the energy separation $\hbar
\Omega$ of the two level system. 

The energy distribution is certainly far from being Gaussian. 
Thus the mean square deviations given by Eqs. (\ref{energyflspin})
and (\ref{energyflosc}) might not be a good indicator of the way the 
energy is distributed over the different states of the system. For instance 
for the two state system discussed above the probability distribution
$P(E)$ to find 
find the system with energy $E$ in the interval $dE$ obviously consists 
of two $delta$-function peaks at $E_{+}$ and  $E_{-}$. Thus we can write 
this distribution in the form, 
\begin{equation}
P(E) = p_{+} \delta(E-E_{+}) + p_{-} \delta(E-E_{-})
\label{pdist}
\end{equation}
where $p_{\pm}$ are as above the probabilities to find the system 
in the excited state and in the lowest energy state of the system. 
Now since $p_{-}+ p_{+} =1$ and since 
$\langle E \rangle \equiv \langle H_s \rangle
 = E_{-} p_{-}+ E_{+}p_{+}$ we can also express the probabilities
$p_{\pm}$ in the form $p_{+} = \langle E \rangle /\hbar\Omega - E_{-}/\hbar
\Omega$
and $p_{-} = E_{+}/\hbar\Omega - \langle E \rangle /\hbar\Omega$. Here we have 
used $\hbar\Omega = E_{+} - E_{-}$. Without loss of generality we 
can set $E_{\pm} = \pm \hbar\Omega/2$ and thus find for the distribution
\cite{jb}
\begin{equation}
P(E) =  \frac{1}{2} (1 + \frac{\langle E \rangle}{2\hbar\Omega})
\delta(E- \frac{\hbar\Omega}{2}) + 
\frac{1}{2} (1 - \frac{\langle E \rangle}{2\hbar\Omega})
\delta(E + \frac{\hbar\Omega}{2})
\label{pdist1}
\end{equation}
Note that if the system and bath are decoupled we have 
$\langle E \rangle = -\hbar\Omega/2$ and the distribution function 
consists of only one peak at $E = - \hbar\Omega/2$. 

We emphasize that 
$\langle E \rangle \equiv \langle H_s \rangle$ 
is the expectation value of the system's energy 
in the overall ground state of the system plus bath plus interaction energy. 
Below we will also discuss $P(E)$ for the harmonic oscillator. For a
harmonic oscillator $P(E)$ consists of an infinite number of 
$delta$-functions with a rapidly decreasing weight of the higher lying
states.

To be specific, consider the two-state system Hamiltonian
to be 
\begin{equation}
H_s= (\epsilon/2)\, \sigma_z +
(\Delta/2)\, \sigma_x
\label{hs}
\end{equation}
where $\epsilon$ measures the distance from resonance 
and $\Delta$ is the energy separation at resonance.
$\sigma_z$ and $\sigma_x$ are Pauli spin matrices. 
The level separation is thus determined by the  
frequency $\Omega=\sqrt{\epsilon^2+\Delta^2}/\hbar$. 
If this system is coupled via $\sigma_z$ to a harmonic oscillator
bath (spin-boson problem) $H_s$ will not commute with the total 
Hamiltonian. For this system 
the probabilities to find the system in the 
excited state and in the low energy state are shown in Fig.
(\ref{qubitprobs}).
At resonance $\epsilon = 0$, the probabilities $p_{\pm}$ tend with 
increasing coupling constant towards $1/2$. For these parameters
we can use the Bethe solution of the anisotropic Kondo model \cite{bethe,cpb}. 
In the strong coupling limit 
an energy measurement will find the system with equal probability 
in the ground state and in the excited state of the system. 
If the 
system is not symmetric $\epsilon \ne  0$ the probability to find the 
system in an excited state reaches a maximum as a function of $\alpha$
and tends to zero for very strong coupling. For large $\epsilon$
the probabilities can be found from perturbation theory \cite{bethe,cpb}. 
Combining both the Bethe Ansatz solutions and the perturbation theory \cite{cpb}
it is possible to give the probabilities 
over the entire range of parameters.
Computational work based 
on renormalization is reported in Ref. \cite{spin-boson}.

Experiments are always carried out at finite temperature,
and it is important to demonstrate that there exists a 
cross-over temperature to the quantum behavior discussed here.
In the low temperature limit, the thermal occupation probability is 
$p_{+} = e^{-\hbar \Omega /k T}$ (see Eq. (\ref{gibbs}).
In the weak coupling limit for the symmetric spin boson
problem, the probability to measure the state as ``spin up''
scales as $p_{+}= \alpha \log (\omega_{c}/\Omega)$. 
Setting these factors equal and solving for $T^{\ast}$ yields \cite{jb}
\be
k T^{\ast} = -\frac{\hbar \Omega}{\log (\alpha \log\frac{\omega_c}{\Omega})}\, .
\label{tcross}
\ee
Since $T^{\ast}$ scales as the inverse logarithm of the
coupling constant, it is experimentally possible to reach
a regime where thermal excitation are negligible.
Experimentally and theoretically \cite{hart} one might be tempted to define
temperature with the help of the qubit by fitting 
$p_{+} = e^{-\hbar \Omega /k T_{eff}}$ with an effective 
temperature to the experimental data. Experimentally 
one would then find that it is impossible to cool the qubit 
below the temperature $T^{*}$. 
Still a careful examination 
would show that the state is not in fact "thermal" since 
it depends on the coupling constant to the bath. 

The temperature $T^{\ast}$ can be viewed as a measure of the 
energy difference between the energy of the lowest energy separable 
state and the true entangled ground state. Recent works
emphasize the role of this energy difference as an entanglement 
witness \cite{dn,ddb}. We now relate energy fluctuations 
directly to entanglement in the ground state.

\section{Energy fluctuations as an entanglement witness}

Energy fluctuations are determined by probabilities
alone, that is by the diagonal
matrix elements of the density matrix only. Thus it is not obvious 
that we can make a statement about entanglement. In general 
such a statement also depends on the non-diagonal 
elements of the density matrix. However, 
we are given the additional information that
we are in the ground state.  
If we measure the subsystem's energy and find
an excited energy, then we know that the state is entangled.
For weak coupling we can make quantitative statement. 
The probability to find the system in higher lying states 
is exponentially
suppressed. To first order in the coupling
constant, we can consider a two-state system where
the density matrix has the form
$p_{-} \equiv \rho_{--}=1-\alpha p$, 
$p_{+} \equiv \rho_{++}=\alpha p$,
$\rho_{+-}=\rho_{-+}^\ast = \alpha c$.
\begin{equation}
\rho =
\left(\begin{array}{cc}
1 & 0  \cr
\noalign{\medskip}
0 & 0
\end{array}\right)
 +  \alpha\,
\left(\begin{array}{cc}
-p & c  \cr
\noalign{\medskip}
c^\ast & p
\end{array}\right) 
+{O}(\alpha^2)\, .
\label{sepd}
\end{equation}
For vanishing coupling constant $\alpha=0$, this just
gives the density matrix for the separable state.
The linear dependence of $\rho$ on $\alpha$ holds
to first order for the model systems considered here 
and is the entanglement contribution.
If one measures the diagonal elements of
$\rho$, one obtains $\rho_{--}$ and
$\rho_{++}$ as the probability to measure the system in
the ground or excited
state. 
If we now diagonalize $\rho$, the eigenvalues
are
$\lambda_\pm = \{ 1- p \, \alpha, p\,  \alpha \} +{O}(\alpha^2)$.
To first order in $\alpha$,
the eigenvalues are the diagonal matrix elements, so
we may (to a good approximation) write entanglement measures
like the purity in terms
of these probabilities. 
As an example the purity $Tr(\rho^{2})$
is given by   $Tr(\rho^{2}) = \lambda_{+}^{2} +\lambda_{-}^{2} \sim
1-2\alpha p = 1-2p_{+}$. 

We next consider two examples of qubits which exhibit 
the behavior discussed above.

\section{Persistent current qubits} 

Thus far we have focused on the energy of the system as the quantity 
of interest. In this section we show that other observables which reflect 
properties of the system behave in fact very similarly. 
In particular we discuss two states systems (qubits) which have the property 
that their tunnel matrix element $\Delta$ in 
Eq. (\ref{hs}) is dependent on an Aharonov-Bohm flux $\Phi$. 
The free energy of such a system then depends on the flux and in 
general supports a persistent current in its ground state 
$I(\Phi) = -dF/d\Phi$. The prediction of persistent currents 
in small normal and disordered loops \cite{bil} has played an important role 
in the development of mesoscopic physics and continuous to be a subject 
of current interest \cite{exp}.

\begin{figure}[t]
\begin{center}
\psfig{file=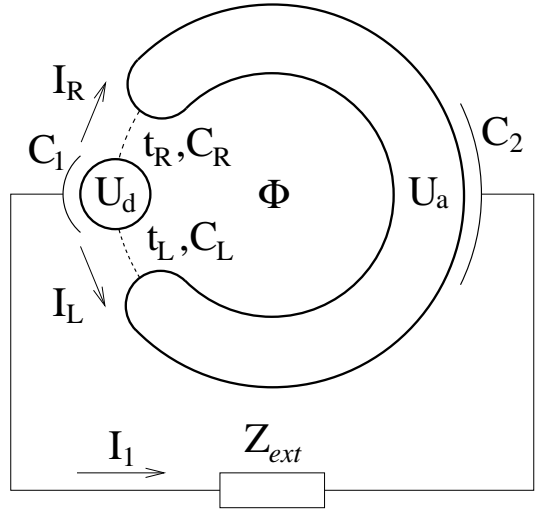,width=7cm}
\caption{Ring with an in-line quantum dot penetrated by an Aharonov-Bohm flux.
The ring is coupled capacitively to an external circuit. After Ref.\protect\cite{cb2}}
\label{ring}
\end{center}
\end{figure}

\subsection{The mesoscopic persistent current qubit} 

A small metallic loop shown in Fig. (\ref{ring}) can be made into a 
two state system with the help of a quantum dot \cite{bs}. 
For sufficiently small charging energy only the states with 
$N$ and $N+1$ electrons on the dot will be relevant. 
As long as these two states are energetically very different,
charge on the dot is fixed and  
transport through the dot is blocked. Only near the 
point of degeneracy can electrons tunnel in and out of the 
dot.  
In the charge basis, the charge on the 
dot is proportional to $\sigma_z$ with an energy  $\epsilon$
that determines how far away the system is from the point 
of degeneracy. At the point of degeneracy,  
charge tunneling is permitted since the dot  
is weakly coupled with 
tunnel energies $t_L$ and $t_R$ to its contacts. 
In such a ring the persistent current exhibits sharp peaks at special values 
of the gate voltage \cite{bs} much like the Coulomb peaks of conductance. 
The Aharonov-Bohm flux can be incorporated into an effective 
tunnel matrix element which connects the charge states $N$ and $N+1$, 
\be
\Delta^{2} /4 = t_L^{2} + t_{R}^{2} \pm 2 t_{L}t_{R} \cos(2\pi \Phi/\Phi_{0}) .
\label{ringdelta}
\ee
Here the sign depends on the number of electrons in the ring. 
Thus near a point of degeneracy the Hamiltonian of this ring is of 
the form given by Eq. (\ref{hs}). We refer the reader to Refs. \cite{bs}
for a detailed derivation. 

The system is coupled to an external circuit  
via capacitances. 
In particular, if the 
external circuit is an ohmic resistor, we can replace it with 
a transmission line with equal input impedance \cite{cpb}. The transmission line 
represents a harmonic oscillator bath \cite{yd} and the entire system 
is a particular realization of the spin-boson problem with the 
interesting feature that the tunnel matrix-element is 
flux dependent. 

The model permits us to address the interesting question
of how persistent currents are affected by environments. 
We follow 
here the discussion of Cedraschi, Ponomarenko and one of the authors \cite{cpb}.
An extended discussion of the initial work is provided in Ref. \cite{cb1}. The
persistent current is obtained using the results of Bethe Ansatz and perturbation 
results for the anisotropic Kondo model \cite{bethe}. For weak system-bath 
coupling, Ref. \cite{cb2} provides a 
discussion based on a quantum Langevin approach. 
Closely related works investigate the effect of a fluctuating Aharonov-Bohm flux \cite{mabr},
the effect of charge fluctuations in nearby pure and 
disordered conductors \cite{guinea,golub}
and the effect of hot bosonic baths \cite{entin}.

Let us now first show that the persistent current, like the energy 
of the system, is not sharp but fluctuates.
We are interested in current fluctuations so we need 
an expression for the current operator. Since the persistent current 
is due to electrons of the ring alone we can consider the isolated system
$\alpha = 0$. Eq.(\ref{hs}) is the Hamiltonian in the charge basis.
In the eigen basis the Hamiltonian is simply $ \hat H_{s} = 
({\hbar\Omega/2}) \sigma_{z}$. In the eigen basis the persistent current 
carried by a state is determined by the derivative of the energy of this state 
with respect to flux. 
The persistent current of the eigenstates
with energies $\pm {\hbar\Omega}/{2}$ is $\mp I_{0}$ with 
$I_{0} = (1/2) d{\hbar\Omega}/d\Phi$.
One of the states
corresponds to a clockwise persistent current and one state corresponds 
to a counter-clockwise current. 
Thus in the eigen basis 
the current operator is simply 
\be 
\hat I = I_{0} \sigma_{z} = - (2I_{0}/{\hbar\Omega}) \hat H_{s} .
\label{perop}
\ee
Here we use that 
in the eigen basis, both the current operator  
and the energy are proportional to $\sigma_z$. 
Thus the persistent current 
is directly related to the energy of the system \cite{notepc}. 
While the first expression in Eq. (\ref{perop}) is valid only 
in the energy eigen state basis, the second expression is 
in fact general.  

In the presence of a bath,  
the persistent current, like the energy of the two state system, 
is in fact not sharp but fluctuates.  Since the average persistent 
current is 
$\langle \hat I \rangle  = I_{+} p_{+} + I_{-}p_{-}$
with $p_{+} + p_{-} = 1$ we can also express $p_{\pm}$ in terms of the
persistent current. With $I_{\pm} = \pm I_{0}$ we find 
$p_{\pm} = (1/2) (1 \mp \langle \hat I \rangle /I_{0})$
and the distribution of currents is thus 
\be 
P(I) = \frac{1}{2} (1 + \frac{\langle \hat I \rangle}{I_{0}}) \delta(I - I_{0})
+  \frac{1}{2} (1 - \frac{\langle \hat I \rangle}{I_{0}}) \delta(I + I_{0})
\ee
Note that $p_{\pm} = (1/2) (1 \mp \langle \hat I \rangle /I_{0}) \le 1$
implies that coupling the ring to the bath can only suppress \cite{cpb} the 
persistent current $\langle \hat I \rangle \le I_{0}$.

\begin{figure}[t]
\begin{center}
\psfig{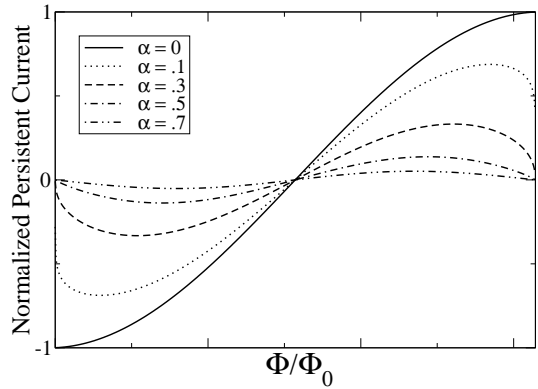}
\caption{Persistent current as a function of the Aharonov-Bohm flux at resonance 
($\epsilon = 0$) for different coupling strengths $\alpha$. Arbitrary small
coupling suppresses the discontinuity of the $\alpha = 0$-persistent current at $\Phi/\Phi_{0} = 0$ and $1$.}
\label{pascal}
\end{center}
\end{figure}

Next let us investigate the flux dependence in more detail. 
In the absence of coupling to a bath $\alpha = 0$, 
and at the point of degeneracy
$\epsilon = 0$ the persistent current is, 
$I_{0}(\Phi) = {d\Delta(\Phi)/d\Phi}$. 
Consider now additionally the special symmetric case, 
when the tunneling rates
are equal $t \equiv t_L = t_R$.
Taking the lower sign in Eq. (\ref{ringdelta})
the persistent current is $I(\Phi ) = -(e/h) 4\pi t \cos(\pi \Phi/\Phi_0 )$
in the interval $0 \leq \Phi \leq \Phi_{0}$. It is a periodic 
function with a discontinuous jump at $\Phi = n \Phi_{0}$, $n = 0, \pm 1, \pm
2,..$. 
It is shown as a solid line in Fig. (\ref{pascal}). 

It is now very interesting to investigate what happens to the discontinuous 
jump in the persistent current in the presence of the bath. Since 
a Fourier representation of the persistent current
\be
I(\Phi) = \sum_{n} I_{n} \sin(2\pi n \Phi/\Phi_{0})
\ee
needs arbitrary high 
harmonics such a jump is a signature of a perfectly 
coherent system: the top-most electron in our system which gives rise 
to this current must circulate the Aharonov-Bohm flux coherently 
$n$-times to generate the $n-th$ harmonic.  
Coupling to a bath
generates de-coherence, and we suspect that the bath 
suppresses such a discontinuity
immediately. The Bethe Ansatz solution of the spin-boson model
gives a
persistent current 
\be
I(\Phi) \propto \Delta^{(\frac{\alpha}{1-\alpha})} d\Delta/d\Phi
\label{perbethe}
\ee   
where $\alpha$ is the coupling constant. The exact result is given 
by Cedraschi et al. \cite{cpb}. The persistent current as a function of 
flux for different coupling constants is shown in 
Fig. (\ref{pascal}). Thus an arbitrary 
small coupling to a bath is sufficient to 
suppress the discontinuity in the persistent current.  
Using  Eq. (\ref{perbethe})
a more quantitative analysis can be provided. Pilgram \cite{sp} finds 
for the Fourier amplitudes 
\be
\frac{I_{n}}{I_{1}} = 
n \frac{(2\alpha -1)...((2n-2) \alpha -(2n-3)}
{(4\alpha - 5).....(2n\alpha -(2n+1))}
\ee 
With the Ansatz 
\be
\frac{I_{n}}{I_{1}} = A_{n} \exp(-b_{n}(\alpha ) \alpha (n -1))
\ee 
the values found for $b_{n}(\alpha)$ are, $b_{2} = 6/5$,  $b_{3} = 88/105$,
$b_{4} = 626/945$... Thus the bath suppresses the persistent current almost 
in an exponential manner, as if the system were subject to dephasing. This 
suppression is the stronger the higher the harmonics. 

The system of Fig. (\ref{ring}) can viewed as a double quantum dot. 
Recent experimental work by Hayashi et al. \cite{hayashi} has demonstrated that 
conditions can be achieved for which the double dot system is well 
described by a Hamiltonian of the form Eq. (\ref{hs}). The work also 
demonstrates how strong measurements can be implemented. Refs. \cite{aguado,lehur}
represent closely related theoretical work. 

\begin{figure}[t]
\begin{center}
\psfig{file=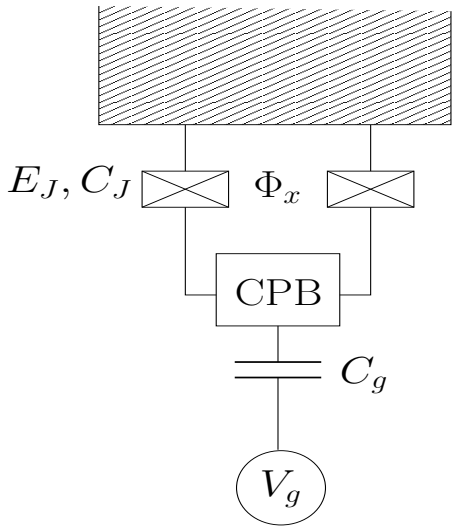,width=6cm,height=7cm}
\caption{Cooper pair box: A superconducting metallic dot is coupled via 
two Josephson junctions (with energy $E_J$ and capacitance $C_J$) to a 
superconductor terminal and is capacitively coupled (with capacitance $C_g$)
to a gate permitting the control of charge. An external flux $\Phi_x$
through the hole of the structure controls the Josephson energy.}
\label{cpbfig}
\end{center}
\end{figure}

\subsection{The split Cooper pair box qubit}

In this section we compare briefly a 
superconducting structure \cite{cpbox1,cpbox2,cottet,qubit}
with a behavior that is analogous to the model discussed above. 
This example is important since, in contrast to the normal state
qubit discussed above, it does not depend on single 
particle energies and since 
a projective measurement testing the state of the system 
by measuring its persistent current, has in fact been implemented \cite{qubit}.
A review of the current research on superconducting 
qubits is provided by Devoret, Walraff and Martinis \cite{devoret}.
The structure of interest here is a Cooper pair box \cite{mb87}, 
a small superconducting island 
coupled with tunnel junctions to a large superconductor and 
coupled capacitively to a gate  (see Fig. \ref{cpbfig}). 
The small superconducting island 
can be split \cite{cottet} such that it forms 
together with the large superconductor a ring. 
If the capacitances of the junctions $C_{J}$ and the capacitance to the gate 
$C_{g}$ are small, the energy for charging the island with an additional 
Copper pair $E_{C} = (2e)^{2}/(2C_{J}+C_{g})$ is large and the system can 
effectively be described with a two-state Hamiltonian. The limit of interest 
is the charge controlled Cooper pair box \cite{devoret} 
for which the Josephson energy 
$E_J$ is much smaller than the charging energy $E_{C}$. The parameters
\cite{devoret} of 
the two-state Hamiltonian Eq. (\ref{hs}) are
$
\epsilon = E_{J} \cos(\pi\Phi_{x}/\Phi_{0})
$
where $\Phi_{x}$ is the externally applied flux, $\Phi_{0}= h/2e$ 
the charge 2e-flux quantum and 
$
\Delta = E_C (1/2 - N_g )
$
where $N_g$ can be controlled by adjusting a gate voltage. 
Comparing the superconducting qubit to the normal conducting qubit 
we notice that here it is $\epsilon$ which is flux dependent. 

We can proceed with the discussion of the persistent current as 
in the normal case: In the eigen basis of the Hamiltonian 
the persistent current in the low and higher energy states 
is $\pm I_{0} = \pm d\hbar \Omega/d\Phi$ with 
$\hbar \Omega / 2 = 
[E^{2}_{J} \cos^{2}(\pi\Phi_{x}/\Phi_{0}) + E^{2}_C (1/2 - N_g )^{2}]^{1/2}$
The operator of the persistent current is as for the mesoscopic qubit given by
$\hat I = - (2I_{0}/{\hbar\Omega}) \hat H_{s}$. The persistent 
current of this superconducting qubit, when it is coupled to a bath (external circuit)
fluctuates even in the ground state. In the experiment of Vion et al. \cite{qubit} 
Rabi oscillations 
are reported with an extremely small damping, a $Q$-factor as high as  
25'000. This value can be used to estimate the coupling constant $\alpha$ and 
leads to a probability to find the system in the excited state of only 
$10^{-4}$. This is to small to be of significance when compared to other effects, 
due the measurement circuit or nearby charge traps. The effect we have discussed 
is most relevant in the strong coupling case between system and bath. Recent  
experiments in which the Cooper pair box is coupled to a cavity 
mode \cite{schoel} offer ways to explore the strong coupling limit of interest here.

\section{Energy fluctuations of the harmonic oscillator}

\begin{figure}[b]
\begin{center}
\leavevmode
\psfig{file=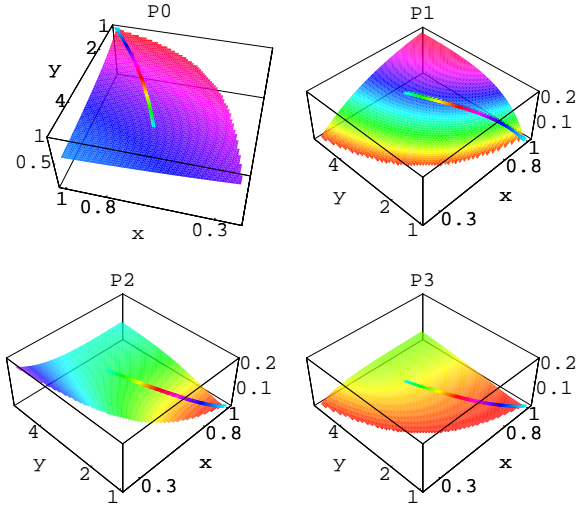,width=8cm}
\caption{The probability to measure a harmonic
oscillator in the ground and first three excited states
as a function of $x$ and $y$ (see text).
The line traces out the behavior of the ohmic bath
as a function of the coupling in the under-damped range.
After \protect\cite{jb}}
\label{bathexcite}
\end{center}
\end{figure}

We now consider the entanglement energetics of a harmonic oscillator,
\be
H_s = p^2/(2 m) + (1/2) m \omega^2 q^2 .
\label{hso}
\ee
Since there are an infinite number of states, the problem
is harder. We assume a linear coupling
with a harmonic oscillator bath.
As a consequence the density matrix is Gaussian so that
environmental information is contained in the second 
moments $\langle q^2 \rangle $ and $\langle p^2 \rangle$ \cite{weiss},
\be
\langle q \vert {\rho}\vert q' \rangle = \frac{1}{\sqrt{2 \pi
    \langle q^2 \rangle}} \exp\left\{-\frac{(\frac{q+q'}{2})^2}
{2\langle q^2 \rangle} -  \frac{\langle p^2 \rangle
  (q-q')^2}{2\hbar^2}\right\}.
\label{dm}
\ee
Expectation values of higher powers of $H_s$ are non-trivial because $q$ and
$p$ do not commute.
The purity of the density matrix Eq.~(\ref{dm}) is
\be
{\rm Tr} \rho^2 = \int dq dq' \langle q \vert \rho \vert q'\rangle
\langle q' \vert \rho \vert q\rangle = \frac{\hbar/2}{\sqrt{\langle
    q^2\rangle \langle  p^2\rangle}} \, .
\label{pure}
\ee
The uncertainty relation, 
$\sqrt{\langle q^2\rangle \langle  p^2\rangle} \ge \hbar/2$,
guarantees that ${\rm Tr} \rho^2 \le 1$ with the
inequality becoming sharp if the oscillator is isolated from the
environment. As the environment causes greater deviation from the Planck
scale limit, the state loses purity.

\subsection{Energy cumulants}

We calculate the generating function 
$Z(\chi)=\langle \exp (-\chi H_s)\rangle$ 
from which we can determine the 
n$^{th}$ energy cumulant
\begin{equation}
\langle\langle H_s^n \rangle \rangle
 = (-)^n \frac{d^n}{d \chi^n} \ln Z(\chi)\Big\vert_{\chi=0}\; .
\label{cumdef}
\end{equation}
by taking derivatives. Ref. \cite{jb} finds 
\be
Z = \left \{ 2 E \,\frac{\sinh \varepsilon\chi}{\varepsilon}
+2 A \,(\cosh \varepsilon \chi -1) + \frac{1+ \cosh \varepsilon \chi}{2}
\right\}^{-\frac{1}{2}}
\label{zans}
\ee
where
$\varepsilon = \hbar \omega$,
$2 E = m \omega^2 \langle q^2\rangle + \langle p^2\rangle /m$ and
$A =  \langle q^2\rangle \langle p^2\rangle /\hbar^2$.
$E$ is the average energy of the oscillator, while
$A\ge 1$ is a measure of satisfaction of the uncertainty principle.

Using Eq.~(\ref{cumdef}),
the first few harmonic oscillator energy cumulants are
straightforwardly found via Eq.~(\ref{cumdef}),
\begin{eqnarray}
\langle\langle H_s^2\rangle\rangle 
&=&(1/2)[-(\varepsilon^2/2) + 4 E^2
- 2 \varepsilon^2 A]\; , \label{2cum}
\\
\langle\langle H_s^3\rangle\rangle  
&=&-(E/2)[-16 E^2 + \varepsilon^2(1+12 A)]\; ,
\\
\langle\langle H_s^4 \rangle\rangle
&=&48 E^4 - 4  \varepsilon^2 E^2 (1+12 A) \nonumber \\
&+& \varepsilon^4 [(1/8) + 2 A + 6 A^2] \, .
\label{cums}
\end{eqnarray}
After inserting the mean square values for an ohmic bath
(see the discussion above Eqs.~(\ref{x},\ref{y})),
Eq.~(\ref{2cum}) is identical to Eq. (\ref{energyflosc}).

\subsection{Density matrix}

Alternatively, we now consider the diagonal matrix
elements $\rho_{nn}$.
An analytical expression for the density matrix in the energy basis
may be found by using the wavefunctions of the harmonic oscillator,
$\psi_n(q) \propto  e^{-\gamma^2 q^2/2} H_n(\gamma q)$
where $\gamma=\sqrt{m \omega/\hbar}$ and $H_n(x)$ is the $n^{th}$ Hermite
polynomial.
In the energy basis, the density matrix is given by
$\rho_{nm} = \int dq dq' \psi^{\ast}_n(q) \langle q\vert \rho \vert 
q' \rangle  \psi_m(q')$.
We first define the dimensionless variables
$x = 2 \gamma^2 \langle q^2 \rangle$,
$y = 2\langle p^2 \rangle/( \gamma^2 \hbar^2)$, and
$D= 1+x+y+x y$.
$x$ and $y$ are related to the major and minor axes of an uncertainty ellipse.
The isolated harmonic oscillator (in it's ground state) obeys two
important properties:  minimum uncertainty (in position and momentum)
and equipartition of energy between average kinetic and potential energies.
The influence of the environment causes deviations from these ideal behaviors
which may be accounted for by introducing two new parameters,
$a=(y-x)/D,\; b=(x y -1)/D$ with  $-1 \le a\le 1$ 
and $0 \le b\le 1$. The deviation from
equipartition of energy is measured by $a$, while the deviation from
the ideal uncertainty relation is measured by $b$.
We find
\be
\rho_{nn} = \sqrt{\frac{4}{D}} (b^2-a^2)^{n/2} P_n\left[b/\sqrt{b^2-a^2}\right]\; ,
\label{pnn}
\ee
where $P_n[z]$ are the Legendre polynomials.
The probability for the lone oscillator
to be measured in an excited state clearly decays rapidly with
level number.
These probabilities also reveal environmental information.
For example, $P_1 =2 b$ and is thus
only sensitive to the area of the state, while
$P_2 = a^2+ 2 b^2$ depends on both the uncertainty and energy
asymmetry. 
Additionally, if we expand the first density matrix
eigenvalue \cite{weiss} with respect to small deviations of
$x$ and $y$, we recover $\rho_{11}$ in agreement with
our general argument.  
%

Thus far we have treated $x$ and $y$ as independent variables. 
In reality the environment the system is coupled to replaces these
variables with two functions of the
coupling constant.  For example, with the ohmic bath \cite{weiss,nb}
(in the under-damped limit), the variables are
\begin{eqnarray}
&& x(\alpha) = \frac{1}{\sqrt{1-\alpha^2}} 
\left(1-\frac{2}{\pi}{\rm \arctan}
\frac{\alpha}{\sqrt{1-\alpha^2}}\right)\; ,
\label{x} \\
&& y(\alpha) = (1- 2 \alpha^2) x(\alpha) + \frac{4 \alpha}{\pi} 
\log\frac{\omega_c}{\omega}\; ,
\label{y}
\end{eqnarray}
where $\alpha$ is the coupling to the environment
in units of the oscillator frequency and $\omega_c$ is a high frequency cutoff.
This bath information is shown in Fig.~(\ref{bathexcite})
with $\omega_c = 10 \omega$.
The  trajectory of the
line over the surface shows how the probabilities evolve as the coupling
$\alpha$ is increased from 0 to 1.  Other kinds of environments would trace
out different contours on the probability surface.

\begin{figure}[t]
\begin{center}
\psfig{file=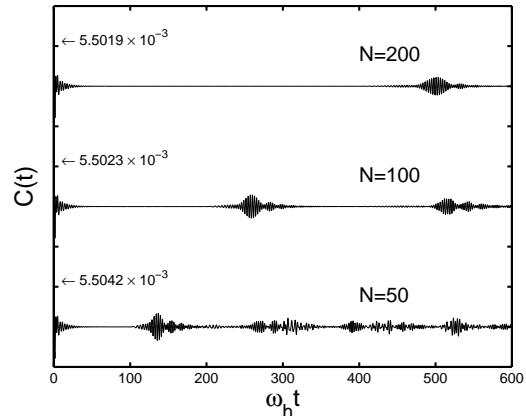,width=7cm}
\caption{Energy-energy correlation 
of a harmonic oscillator coupled to a chain 
of $N$ particles as a function of time. 
The correlation is evaluated in the 
ground state of the entire system. 
The length of the chain is $N =50$ (bottom), $N =100$ (middle), 
and $N =200$ (top).
After Ref. \protect\cite{nb}.}
\label{nagaev}
\end{center}
\end{figure}

\subsection{Ground state energy-energy correlations}

The expectation value of observables of the system, like the energy $\hat H_{s}$, 
the persistent current $\hat I$ or their moments, are 
time-independent in the ground 
state of the total system. However, this is not true for two-time correlations,
like the energy-energy correlation function, 
\be
C(t) \equiv \frac{1}{2} \langle \Delta \hat H_{s}(t) \Delta \hat H_{s}(0) + \Delta \hat H_{s}(0) \Delta \hat H_{s}(t)\rangle .
\ee
Here $\Delta \hat H_{s}(t)= \hat H_{s}(t) -\langle \hat H_{s}\rangle $ are the energy fluctuations 
away from the average energy. This correlation function vanishes
if the ground state is a product of a system and bath wave function. The correlation 
is thus also a measurement of the degree of entanglement between system and bath. 

For the oscillator, Eq. (\ref{hso}), coupled 
to a linear chain of $N$ particles with elongation $x_n , n=1, ..,N$ coupled 
with an energy $(1/2)m_h \omega^{2}_h (x_{n-1} - x_{n})^{2}$ this correlation 
function was calculated by K. E. Nagaev and one of the authors \cite{nb}. 
An infinitely 
long chain generates friction proportional to $\eta = (m_{h} /m )\omega_h$
giving rise to a system bath coupling constant 
$\alpha = (m_{h} /m) (\omega_h /\omega )$.
The calculation 
proceeds by first searching the normal modes of the classical problem. 
The corresponding 
classical problem is then quantized. The energy of the subsystem and 
in particular the energy-energy correlation is written in terms of 
the normal modes of the entire system. The correlation is shown in 
Fig. (\ref{nagaev}) as a function of time for three different
chain lengths  $N = 50, 100, 200$ with $\omega /\omega_{h} =1$ and 
$m_{h}/m = 0.1$. The numbers on the vertical axis 
indicate the 
initial value of the correlation. There is a very rapid initial 
decay of the correlation followed by 
a damped oscillatory behavior. Since the chain is of finite length 
a partial revival is seen after a time 
it takes a perturbation to travel down the chain and back to the 
oscillator. We emphasize that the revival is not complete. The 
spectrum of the chain alone 
would consists of commensurate frequencies, however, due to the 
presence 
of the harmonic oscillator the spectrum of the entire system is 
not commensurate. 

Unlike in the case of the two-state systems we can not easily develop
a model for the harmonic oscillator which connects it to persistent 
currents. However, the conductance in an 
Aharonov-Bohm geometry in which electrons traverse
an oscillatory potential  has recently been discussed by 
Ratchov et al. \cite{hekking}.  
   
\section{Discussion}

The energy of the subsystem is an 
observable which illustrates best the distinction
between separable and entangled ground states. 
We have shown that projective measurements
of the system Hamiltonian at zero temperature
can find the system in higher energy states.
This is the case if the many-body quantum mechanical ground state
of system and environment are entangled.

Entanglement assures that the system "knows" about its
environment and similarly there is system information
in the environment. If the environment is represented as a linear
chain of particles the system bath interaction can be viewed as
a scattering problem. System information is 
then present in the bath in the form of the phase of 
the reflected part of the scattering state \cite{geilo}.

We emphasize that the system bath interaction is not just a
question of renormalization. For instance in the two state problem
discussed here the tunnel matrix element $\Delta$ is "renormalized"
to $\Delta_{eff}$. However, clearly the physics we have discussed
here is not captured by simply replacing in the two-state Hamiltonian
the tunnel matrix element by its renormalized value. Both a two-state
system or a harmonic oscillator 
(with renormalized mass and frequency)
would exhibit a sharp energy in their ground state.

One natural question is how this work is connected to the presence or
absence of dephasing at zero temperature.  Historically, dephasing has
mainly dealt with the randomizing of a quantum mechanical phase via
interaction with some fluctuating variable, such as a reduction of the
Aharanov-Bohm interference pattern from voltage fluctuations, which
usually freeze out at low temperature.  Typically, the off-diagonal
density matrix elements decay in time, while the diagonal elements
stay constant.

A more modern view is to call decoherence any mechanism where one
starts with a pure system state, and ends with a mixed system state.
The ultimate cause of this process is simply entanglement of the
system under observation with other degrees of freedom that are
unmonitored.  Thus, although the entire quantum system may be in a
pure state, the fact that local measurements on the subsystem extract
only part of the information, results in mixed behavior.

In this sense, there is trivially decoherence at zero temperature,
unless either the coupling constant vanishes (so the ground state is
separable), or it is possible to make measurements on every coupled
quantum degree of freedom, so the purity of the many-body ground
state is accessible.

Relaxation into 
equilibrium is probably the simplest possible preparation 
method of an entangled state. For this reason ground state 
entanglement energetics 
will likely be an important direction of future research.

This work was supported by the Swiss NSF
and by the network MaNEP.


\begin{thebibliography}{9}
\bibitem{notepc}
Depending on whether particle currents or total currents 
(including the displacement current) are considered 
different expressions for the persistent current 
operator can be found in Refs.  \cite{bs,cpb,cb1,cb2}. 
For the symmetric case, equal tunnel amplitudes and 
capacitances and $\epsilon = 0$ all these expressions 
are identical.
\end{thebibliography}
\end{document}